# Eight Physicists and Astronomers: Biographical Portraits

Helge Kragh[*]

**Abstract**  This essay provides concise biographical information about eight physicists, astronomers, astrophysicists, and cosmologists from the twentieth century. The portrayed scientists are Hermann Bondi (1919-2005), Charles L. Critchfield (1910-1994), Arthur E. Haas (1884-1941), Chushiro Hayashi (1920-2010), Gustave Le Bon (1841-1931), Wilhelm Lenz (1888-1957), Franz Selety (1893-1933?), and S. Elis Strömgren (1870-1947). Because the entries were written for a biographical dictionary of astronomers, the emphasis is on the astronomical and astrophysical contributions of the scientists rather than their work in physics. While some of them are well known, others (such as Selety, Haas and Le Bon) will probably be unknown to most physicists.

## Bondi, Hermann

*born*        Vienna, Austria, 1 November 1919.
*died*        Cambridge, England, 10 September 2005.

Hermann Bondi, an eminent mathematical physicist, astronomer, and cosmologist, is probably best known as one of the founders of the steady state theory of the universe, which he defended between 1948 and the mid-1960s. In addition to his work in cosmology, he also contributed to a variety of problems in astrophysics and general relativity theory, including gravitational radiation and the accretion of matter by stars in a gas cloud.

Bondi's father was a non-observant Jew who practiced as a doctor, and his mother came from a prosperous Jewish family. After attending a secondary school in Vienna, where he showed an early aptitude for mathematics, in 1937 Bondi succeeded in being admitted as a foreign student at Trinity College, Cambridge. Being an Austrian citizen and thus an "alien," he was interned in May 1940 and spent about a year in an internment camp in Quebec, Canada. While interned, he met another Austrian "alien," Thomas Gold, with whom he would eventually collaborate in developing the steady state theory. Upon his return to England he continued his studies in theoretical physics and was drawn into military research in a group headed by Fred Hoyle. In 1943 Bondi and Gold rented a small house and began regularly to discuss scientific questions with Hoyle. In 1948 Bondi was appointed university lecturer in Cambridge. By that time he had become a British subject and married Christine Stockman, an astrophysics research student with whom he had five children.

[*] Centre for Science Studies, Aarhus University, Denmark. E-mail: helge.kragh@ivs.au.dk. The biographical entries are contributions to Thomas Hockey et al., eds., *The Biographical Encyclopedia of Astronomers*, to be published by Springer.

Except for extended travels to the United States and elsewhere, he stayed in England throughout his life, from about 1970 sharing his work between research and an increasing engagement in science policy and organization. In 1967-1971 he was director general for the European Space Research Organization, ESRO (the predecessor of ESA, the European Space Agency) and later chief scientific advisor for the Ministry of Defense and chairman of the Natural Environment Research Council. Among his many honours were the Einstein Society Gold Medal (1983), the Planetary Award (1993), and the Gold Medal of the Royal Astronomical Society (2001). He was appointed Knight Commander of the Bath in 1973.

One of Bondi's first scientific papers was a masterly review report on cosmology, written on the request of the Royal Astronomical Society and published in early 1948. Although the report did not include the steady state theory, it reflected the cosmological discussions he had at the time with his friends Hoyle and Gold. Four years later, in 1952, he extended the review into a small book, entitled *Cosmology*, which was one of the first books on the subject and widely used as a textbook. While the book systematically covered relativistic evolution theories, and also the cosmological system of E. Arthur Milne known as "kinematic relativity," it also included a detailed account of his own favourite cosmological model, the steady state theory of the universe. This theory was presented in two versions in the early fall of 1948, one by Hoyle and another by Bondi and Gold. Both versions had in common the assumption of an unchanging, yet expanding universe from which followed that matter must be created continually throughout the universe (in order to secure a mass density independent of time). The Bondi-Gold version was explicitly based on what they called the "perfect cosmological principle," the postulate that the universe is homogeneous in both space and time. They argued that this postulate should have priority over the principle of energy conservation, and that without it cosmology could not be a proper science.

The new steady state cosmology aroused much controversy for about fifteen years, in particular in England. During the heated debate, Bondi defended the theory skillfully, arguing that it was superior to relativistic evolution theories from both a methodological and observational point of view. At one stage he suggested, together with the Cambridge astronomer Raymond Arthur Lyttleton, an "electrical cosmology" of the steady state type that presumed a tiny numerical charge difference between the proton and the electron. However, this cosmological model soon turned out to be untenable. The theory of Hoyle, Bondi and Gold ran into serious troubles in the early 1960s, when radio-astronomical data were shown to disagree with its predictions, but for a while Bondi continued to defend it. Then, in 1965 the discovery of the cosmic microwave background radiation made the classical steady state theory impossible. Rather than accepting what some years later became the standard hot big bang cosmology, he chose to leave cosmology as a research field and focus on non-cosmological aspects of astrophysics and general relativity theory.

These subjects were not new to Bondi. For example, starting in the mid-1940s he did very important work on accretion theory, demonstrating that accretion of interstellar matter played a much more important role in the evolution of stars than previously estimated. In 1952 he collaborated with Edwin Ernest Salpeter in a study of thermonuclear reactions in stars, but this line of research he did not follow up. His first work in general relativity was a study of 1947 of the solution of Einstein's cosmological field equations in the case of a spherically symmetric, inhomogeneous universe. The solution is today known as the Tolman-Bondi model because the problem had been considered earlier by Richard Chase Tolman, in a paper of 1934 (and even earlier by Georges Édouard Lemaître). Bondi's main work in general relativity was however gravitational radiation, a subject he examined in an extended series of papers between 1959 and 2004. In 1954 he had become professor of applied mathematics at King's College, London, where he established an important research group in general relativity, a branch of physics which at the time attracted renewed interest. Gravitational radiation was a focal area of research in Bondi's group, among the members of which were his collaborators Felix Pirani and Ivor Robinson. He succeeded to show conclusively that the general theory of relativity predicts gravitational waves and that these should be observable. In the case of plane gravitational waves, a subject he studied in great detail, he proved that they transport energy and thus are physically real.

Bondi had wide-ranging interests, including education, philosophy and politics. He was a friend and great admirer of Karl Popper, whose "falsificationist" philosophy of science he considered the ideal that any science should strive towards. Although of Jewish descent, he never felt attracted to Mosaic faith or any other religion. On the contrary, he was an atheist and often spoke out on behalf of atheism or "humanism," which he thought was congruent with the true spirit of science.

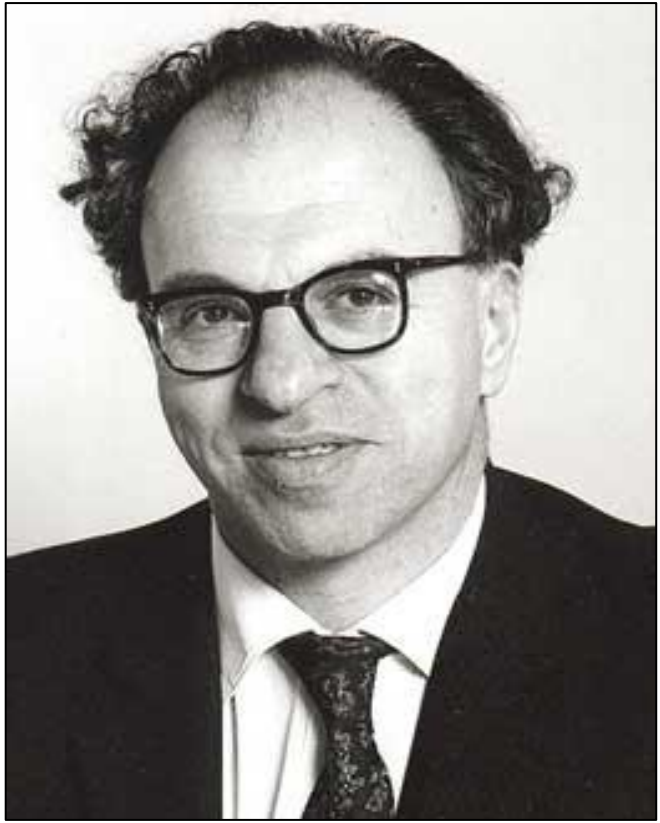

Hermann Bondi

## Critchfield, Charles Louis

*born*        USA, 1910
*died*         Los Alamos, USA, 12 February 1994

American theoretical physicist Charles Critchfield is known in the history of astrophysics for a pioneering joint work with Hans Albrecht Bethe on stellar energy production. He did not follow up on this work or other aspects of nuclear astrophysics. Instead he shared his career between military research, mathematical physics, and theoretical studies in nuclear and particle physics.

Critchfield studied physics at the George Washington University, Washington D.C., where his graduate supervisors were George Antonovich Gamow and Edward Teller. His thesis of 1939 was on strong couplings in particle physics, a subject on which he later published papers with Teller and Eugene Wigner. Between 1939 and 1942 he had positions at various institutions, including the Department of Terrestrial Magnetism of the Carnegie Institute. There he got involved in defense-related work, and in April 1943 he joined the Manhattan Project. In Los Alamos he worked on the gun that, in the case of the uranium bomb, should unite two subcritical masses into one supercritical and explosive mass. After a brief stay at Oak Ridge, in 1947 Critchfield was appointed professor of physics at the University of Minnesota. During a sabbatical leave 1952-1953 he worked on the hydrogen bomb project. After five years in industry, working for General Dynamic's Convair Division in San Diego, he returned to Los Alamos where he taught and did research until his retirement in 1977.

In 1938, while still a graduate student, Critchfield studied the energy liberated in nuclear processes starting with proton-proton collisions and culminating in the synthesis of helium. At a conference organized by Gamow in Washington D.C. the same year, Critchfield met Bethe, and the two collaborated to develop the scheme into a theory for stellar energy production. Their joint paper of August 1938 assumed proton-proton reactions into deuterons to be the fundamental process in the stars. After deuterons had been formed, they would react with protons in a chain leading to helium. The Bethe-Critchfield or "proton-proton cycle" was the first quantitative theory of energy production in stars, and it agreed remarkably well with observations. Using the best estimates for the hydrogen content and temperature in the interior of the sun, they calculated a value for the energy flux close to the one measured. This work led directly to Bethe's famous theory of the CNO cycle published in 1939.

Critchfield collaborated on two occasions with Gamow, first in 1939 on a detailed study of the shell-source model of stars that Gamow had proposed and which he had much confidence in. In 1947 they coauthored the third edition of a book Gamow had written on nuclear physics. The Gamow-Critchfield book contained detailed chapters on nuclear processes in the stars and also a chapter on the origin of the elements in the earliest phase of the expanding big bang universe.

## Haas, Arthur Erich

*Born*        Brno, Czech Republic, 30 April 1884
*Died*        Chicago, USA, 20 February 1941

Austrian-American physicist Arthur Erich Haas made contributions to atomic and quantum theory, and he was also a pioneer in the historical study of the physical sciences. Since the late 1920s he engaged in theoretical work on cosmology and astrophysics, arguing that the expanding universe is finite in both space and time.

The oldest son of Gustav Haas, a lawyer, Arthur Haas grew up in Brno, Moravia, which at the time was part of the Austrian-Hungarian empire. In 1924 he married Emma Beatrice Huber, with whom he had two sons. He studied at the universities in Vienna and Göttingen, and in 1906 he received his doctorate from the University of Vienna, his dissertation being on ancient theories of light. His postgraduate studies included atomic theory as well as history and philosophy of science, which in 1909 resulted in an important monograph on the history of the principle of energy conservation. The following year he proposed for the first time an atomic model incorporating Planck's constant of action. On the basis of his model for the hydrogen atom he derived an expression for the Rydberg constant in terms of elementary constants, of the same kind as the one Niels Bohr presented three years later. However, while Bohr argued that Planck's constant was irreducible, Haas expressed it in terms of the mass and radius of the hydrogen atom. Having been appointed professor of history of science at Leipzig in 1913, he gave courses in theoretical physics and published in 1919 a widely used textbook, *Einführung in die theoretische Physik* (English translation 1925). In 1920 he developed a theory of the isotope shift in band spectra similar to the one that independently was proposed by Adolf Kratzer in Germany and Francis Wheeler Loomis in the United States. From 1923 to 1935 he was professor at the University of Vienna, during which period he wrote several textbooks, including an introduction to wave mechanics (1928) and the first book on quantum chemistry (1929). In 1935 he emigrated with his family to the United States, where he first stayed as a visiting professor at Bowdoin College, Maine. With the support of Albert Einstein, the next year he joined the faculty at the University of Notre Dame. He remained at Notre Dame until his death in 1941. Among other activities, in 1938 he organized an important meeting on "The Physics of the Universe and the Nature of Primordial Particles," one of the very first conferences on physical cosmology. In this meeting participated, among others, Harlow Shapley, Arthur Holly Compton, Georges Édouard Lemaître, and Carl David Anderson.

Haas' interest in cosmological questions was in part of a philosophical and theological nature, as shown by papers of 1907 and 1911 in which he argued that an eternal universe was inconsistent with the laws of physics. He based the conclusion not only on the second law of thermodynamics but also on the finite life-times of radioactive elements. This was the first time radioactivity was considered in a cosmological context. Much later, and especially after he had settled in the United States, cosmology became his main area of research. In a book of 1934, *Kosmologische Probleme der Physik*, he gave one of the first introductions to the new relativistic cosmology of the expanding universe. In part inspired by the ideas of Arthur Stanley Eddington, he suggested various relations between atomic and cosmic constants of nature, including an expression of the mass of the universe in terms Hubble's expansion constant and the classical electron radius. Like Lemaître, he always insisted that the universe must be spatially finite, and in 1938 he suggested that its total energy might be zero, meaning that the positive mass-energy was compensated for by its negative gravitational energy. This idea was later taken up by other cosmologists, including Pascual Jordan and Charles W. Misner. Haas was a Catholic, and in a paper of 1938 he addressed the spiritual meaning of cosmology and its relationship to theism. He considered the closed finite-age universe to be consistent with his Christian belief, while he argued that it was inconsistent with atheism. Although he welcomed Lemaître's "primeval atom" universe – the first version of what came to be known as big bang models – he also stressed that the origin of the universe is a unique process that cannot be understood from the standpoint of physics.

**Hayashi, Chushiro**

*born*          Kyoto, Japan, 25 July 1920
*died*          Kyoto, Japan, 28 February 2010

Chushiro Hayashi (or Hayashi Chushiro) was the most important Japanese astrophysicist in the twentieth century. Based in Kyoto University, he did pioneering work in big bang cosmology, nuclear astrophysics, star formation, and the origin of the solar system. He was among the first to make extensive use of electronic computers in astrophysical calculations. A leader of Japanese astrophysics, he formed a strong research school and had many graduate students.

    After graduating high school in Kyoto, Hayashi studied physics at the University of Tokyo. He graduated in 1942 and was then conscripted into the Navy. In 1945, when the war had ended, he returned to Kyoto to resume his studies in particle physics and astrophysics under Hideki Yukawa, Japan's first Nobel laureate. His doctoral degree of 1954 was based on a thesis on quantum field theory, but he subsequently changed to astrophysics. In 1957 he was appointed professor at Kyoto University, where he stayed until his retirement in 1984. Hayashi was honoured with many prizes,

including the Eddington Medal (1970), the Kyoto Prize (1995), and the Bruce Medal (2004).

According to the hot big bang model proposed in 1948 by George Gamow, Ralph Alpher and Robert Herman, the universe initially consisted of neutrons. In a paper of 1950, Hayashi, then at Nanikawa University, argued that effects of the neutrinos had to be taken into account and that this would result in an original state with more protons than neutrons. He calculated a present hydrogen-to-helium ratio of the right order of magnitude, which inspired Alpher and Herman to improve their big bang model. In a follow-up paper of 1956 Hayashi and his collaborator Minoru Nishida suggested how heavier elements could be built up from helium in the hot early universe, but the suggestion turned out not to be viable. Hayashi was one of the very few scientists to take the big bang theory seriously before the mid-1960s. In a paper of 1963 he calculated the age of the big bang universe, including a cosmological constant, to be less than 13 billion years.

While Hayashi's contributions to cosmology were not much appreciated, his work in astrophysics won him international recognition. In the 1960s he and his collaborators developed models of star formation and pre-main-sequence stellar evolution. As a by-product of his research on red giant stars, in 1961 he discovered what is known as the "Hayashi phase." He also found that when a star has reached hydrostatic equilibrium, it cannot exceed a certain radius given by the "Hayashi limit." Among other important work done by Hayashi and his group at Kyoto University was an early study of brown stars and computer simulations of how a gas cloud collapses spherically to form a star. During the 1970s and 1980s he was much occupied with studies of the formation of planets and the origin of the solar system.

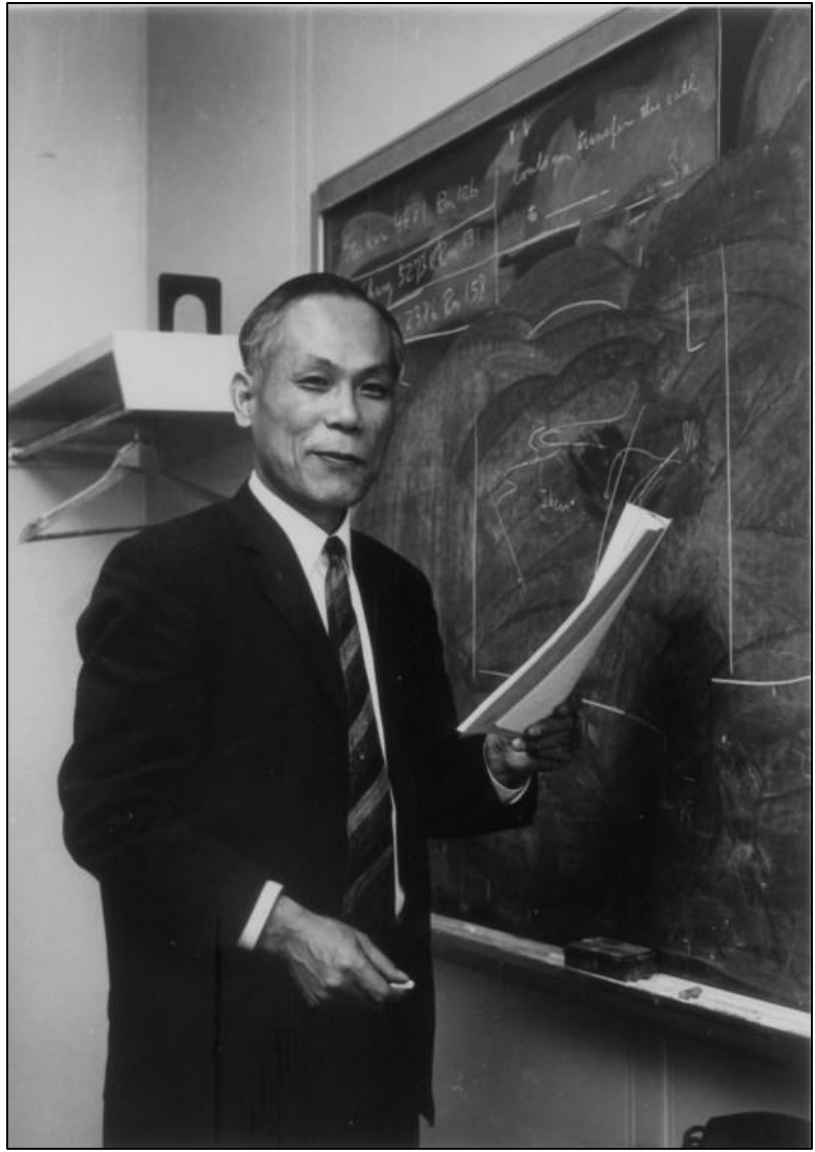

Chushiro Hayashi

## Le Bon, Gustave

*Born*        Nogent-le-Rotrou, France, 7 May 1841
*Died*        Marnes-la-Coquette, France, 13 December 1931

Gustave Le Bon (also spelled LeBon) is primarily known for his influential works in sociology, group psychology, and anthropology, but he also wrote on physical, astronomical, and cosmological subjects. His popular writings on scientific subjects represented evolutionary and antimaterialistic views that held considerable appeal in the fin-de-siècle period, both in France and elsewhere. According to Le Bon, the material universe was born out of a primordial and imponderable ether, and it would eventually return to an ethereal state.

As a young man Le Bon studied medicine in Paris and worked on chemical and physiological experiments in his own laboratory. He made several field trips for sociological and anthropological studies, which he described in papers and books. Inspired by the English philosopher Herbert Spencer, he soon turned to sociology and psychology, in which fields his books were successful and influential. His best-selling work was *La Psychologie des Foules* from 1895, which appeared in an English translation the following year (*The Crowd: A Study of the Popular Mind*). Le Bon's ideas about group psychology were widely discussed and later had some political impact, in particular by being used in fascist theories of leadership. On the other hand, they were criticized by Sigmund Freud, the founder of psychoanalysis. At the same time as he wrote about social psychology, Le Bon made physical experiments and developed a speculative and purely qualitative theory of cosmic evolution and devolution. His main works in this area appeared 1906-1907 in the form of two books that were translated into English as *The Evolution of Matter* (1907) and *The Evolution of Forces* (1908).

In 1896 Le Bon announced that he had discovered what he called "black light," a new kind of invisible radiation that he believed was distinct from, but possibly related to, X-rays and cathode rays. The discovery claim attracted much attention among French scientists, many of whom supported it and Le Bon's general ideas of matter, radiation, and ether. Although black light turned out to be imagination rather than reality – most physicists failed to detect the rays – for a while they established Le Bon as an important figure in French scientific and intellectual life. In 1903 he was even nominated for the Nobel Prize in physics. Among his friends and admirers were several leading scientists, including the chemist Henri Ferdinand-Frédéric Moissan, the astronomer Henri Alexandre Deslandres, and the mathematicians Charles Émile Picard and Jules Henri Poincaré.

More important than the ephemeral black light was Le Bon's cosmo-physical speculations as he described them in *The Evolution of Matter* and elsewhere. He concluded that all matter is unstable and degenerating, constantly emitting rays in the forms of X-rays, radioactivity, and black light. Material qualities were held to be epiphenomena exhibited by matter in the process of transforming into the imponderable and shapeless ether from which it had once originated. Energy and matter were two sides of the same reality, different stages in a grand evolutionary process that in the far future would result in a pure ethereal state. His main argument for the continual degradation of matter into ether was radioactivity, which he considered a property exhibited by all matter. Then, if all chemical elements emitted ether-like radiations, would they not slowly melt away and would this not prove that matter could not be explained in materialistic terms? Indeed, this is what Le Bon concluded. After a period of existence, matter would return to what he called "the final nirvana," that is, the ether. And yet the ether might not be the ultimate end state, for he suggested that the final destruction of the world would perhaps be followed by a new cosmic birth and evolution, and that the cyclic process might go on eternally. Although clearly speculative, many scientists found Le Bon's ideas attractive. For example, the respected British physicist Oliver Joseph Lodge held views that in many ways were similar to that of the French sociologist and amateur physicist.

Le Bon took his version of cosmic evolution from Laplace's nebular hypothesis, but dressed it in the language of the popular ether physics. The primordial ether was the cosmic starting point from which the nebulae originated by condensation. He also argued that all atoms contain enormous amounts of locked energy that would be liberated as the atoms decayed, and that this intra-atomic energy provided the source of solar heat and all other forces in the universe. When Einstein became famous for his demonstration that mass and energy are equivalent according to $E = mc^2$, Le Bon objected that credit belonged to him. In letters of 1922 he informed Einstein of what he considered his priority of having discovered the reciprocity of matter and energy. Of course, by that time his scientific views were completely out of date. While still remembered as a social scientist, he was forgotten as a physicist and cosmologist.

## Lenz, Wilhelm

| | |
|---|---|
| *born* | Frankfurt am Main, Germany, 8 February 1888 |
| *died* | Hamburg, Germany, 30 April 1957 |

Wilhelm Lenz was a leading German physicist who did important work in atomic theory and solid-state physics. His involvement in the astronomical sciences was largely limited to a remarkable paper of 1926 in which he applied thermodynamics and radiation theory to the closed Einstein cosmological model. This was a pioneering contribution to physical cosmology, but it is not well known.

Lenz studied physics and mathematics at the University of Göttingen and subsequently studied under Arnold Sommerfeld in Munich, from where he was granted his doctorate in 1911. For the next nine years he served as Sommerfeld's assistant, only interrupted by military work as a radio operator during World War I. After a brief stay at the University of Rostock, in 1921 he was appointed professor of theoretical physics at the University of Hamburg and director of its institute of theoretical physics. He stayed in this position until his retirement in 1956. At Hamburg, he had a number of brilliant assistants and students, including Wolfgang Ernst Friedrich Pauli, Pascual Jordan, Ernst Ising and Albrecht Otto Johannes Unsöld. The latter, who was another of Sommerfeld's students, specialized in spectroscopic studies of stellar atmospheres and became one of Germany's leading astrophysicists.

Most of Lenz's research was in the area of atomic and quantum physics. In 1919 he proposed improved models of the hydrogen molecule and the helium atom, and in 1924 he analyzed another of the problems of the old quantum theory, the spectrum of hydrogen in the case of crossed electric and magnetic fields. These problems remained anomalous and were only solved after the Bohr-Sommerfeld orbital model had been replaced by the new quantum mechanics. Today Lenz is probably best known for his share in the Ising model, also known as the Lenz-Ising model, which is a general model for atoms arranged on a crystalline lattice and similar cooperative phenomena. Although commonly ascribed to a work done by his student Ernst Ising in 1925, the essence of the model was suggested by Lenz in a study of ferromagnetism from 1920.

Otto Stern, the professor of experimental physics in Hamburg, had in 1925 examined the equilibrium of matter and radiation in the universe and found that even at very high temperatures the content of particles would be exceedingly small. Inspired by Stern's study, Lenz investigated the same problem in the case of the closed and matter-filled model Einstein had proposed in 1917. In his paper of 1926 he argued that the world radius, or curvature of space, depended on the matter-radiation equilibrium. On the other hand, it did not depend on the zero-point radiation energy that Hermann Walther Nernst had derived from quantum theory. Lenz calculated a relation between the radiation temperature $T$ and the world radius, namely $T^2 \cong 10^{31}/R$, where $R$ is the radius in cm. Taking $R \cong 10^{26}$ cm, a value corresponding to the generally accepted mean density of matter $10^{-26}$ g cm$^{-3}$, he arrived at the unrealistically high space temperature 300 K. Lenz's theory was further developed by Richard Chase Tolman and Pascual Jordan, but after the discovery of the expanding universe it fell into oblivion. Only much later, and especially after the discovery of dark energy in the late 1990s, did the gravitational effect of zero-point radiation energy become an issue, and by that time Lenz's work was effectively forgotten.

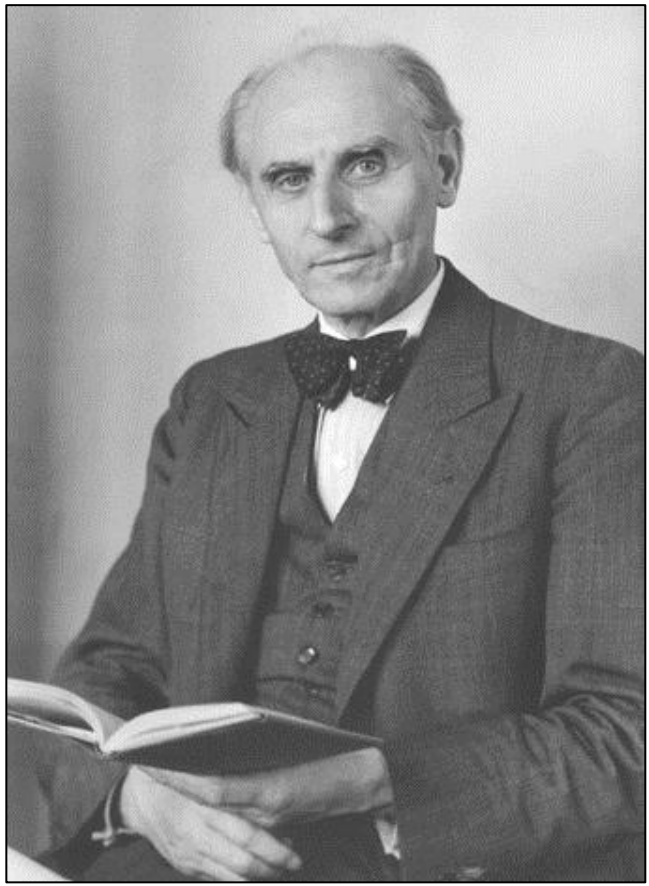

Wilhelm Lenz

**Selety, Franz**

| | |
|---|---|
| *born* | Dresden, Germany, 2 March 1893 |
| *died* | 1933 (?) |

An Austrian citizen, until 1918 the name of Franz Selety was Franz Josef Jeiteles. In the period 1917-1922 he was in contact with Albert Einstein, whose new theory of the universe, based on the general theory of relativity, he criticized in several papers. As an alternative, Selety proposed a Newtonian, hierarchic universe. The model attracted brief attention, but was not further developed. After 1924, no more was heard of either Selety or his cosmological model.

Having completed high school education in Vienna, in 1911 Selety entered Vienna University to study philosophy, at the same time developing an interest in physics, mathematics and cosmology. In 1915 he obtained his doctoral degree in philosophy, and two years later he wrote two long (but no longer extant) letters to Einstein, in which he explained his philosophical views and their relevance for cosmology. A paper on the structure of the universe that he published in *Annalen der Physik* in 1922 caused Einstein to reply. While Selety published three more papers on the subject in 1923-1924, Einstein discontinued the debate. Two of Selety's papers were published in the *Comptes Rendus* of the French Academy of Science and communicated by the mathematician Émile Borel, who had himself an interest in hierarchical cosmologies of the kind proposed by Selety. After 1924, when Selety stayed in Paris, he disappeared from the scene of science and philosophy. He may have died in 1933, but nothing is known of his later years.

Einstein's closed universe of 1917 was homogeneously filled with a finite amount of matter in such a way that it avoided the gravitational paradoxes of Newtonian theory. Selety's model of 1922 developed ideas of an infinite hierarchic (and hence inhomogeneous) universe governed by Newtonian gravitation that the Swedish astronomer Carl Charlier had proposed in 1908. On this basis he showed that, in the case where matter dilutes inversely with the square of the distance, one of Einstein's major arguments in his 1917 theory would be invalid. Contrary to what Einstein had argued, in an otherwise empty universe stars would not disperse but could be held in a stable cluster by its own gravitational force. In his reply of 1922, Einstein conceded his error, but without considering it important. Selety's attempts to involve Einstein in a debate concerning the possibility of a Newtonian rather than a relativistic universe did not bear fruit. Einstein was convinced that the universe was static and finite, and that it could only be described by the equations of general relativity supplemented with a positive cosmological constant.

**Strömgren, Svante Elis**

*born*        Helsingborg, Sweden, 31 May 1870
*died*        Copenhagen, Denmark, 5 April 1947

Swedish-Danish Elis Strömgren was an astronomer of the classical school, all of his scientific work being in areas of astrometry and celestial mechanics. As professor of astronomy in Copenhagen 1907-1940 he had decisive influence on Danish astronomy, and he was also an important figure in international astronomical organizations. His wife since 1902, Hedvig Strömgren, was a dentist, feminist, and successful author of novels. One of their sons, Bengt Georg Daniel Strömgren, became a leading astrophysicist. When Strömgren senior resigned from his position as director of the Copenhagen Observatory in 1940, his son succeeded him.

Elis Strömgren studied astronomy at the University of Lund, Sweden, and in 1898 passed his doctoral degree. He spent the years 1901-1907 in Kiel as lecturer and assistant editor of *Astronomische Nachrichten*. From 1920 until his death in 1947 he was chief editor of *Nordisk Astronomisk*

*Selskab*, a joint astronomical journal for the Nordic countries founded in 1916. Strömgren's main line of research dealt with the original orbits of comets, which he had first investigated in his doctoral dissertation and continued to work on for more than thirty years. His extensive computations backward in time showed that in nearly all cases the orbital elements of near-parabolic comets had changed into a more elliptical direction, indicating that all known comets belong to our solar system. Another line of research to which he devoted much effort was the classical problem of perturbation known as the *problème restraint* ("restricted problem"). While an accomplished astronomer in the mathematical tradition, he had little appreciation of the new astrophysics in which his son specialized.

Strömgren was active in maintaining international relations in astronomy in and after World War I. The Central Bureau of the Astronomische Gesellschaft, located in Kiel, could not continue its activities with distributing astronomical news after 1914, and on Strömgren's suggestion it was moved to Copenhagen. After it had become a department of the new International Astronomical Union (IAU), at the first General Assembly in 1922 it was decided that the bureau should be located in Copenhagen with Strömgren as its director. Representing a country that had stayed neutral during the war, he had good relations to both German, English and French astronomers. His efforts to maintain cooperation and communication between the nations in spite of the war and its aftermath were broadly appreciated, an indication being his nomination for the Nobel Peace Prize in the years 1920, 1922 and 1923. Despite the support of many astronomers, the nominations did not bear fruit. Strömgren was a delegate for Denmark at the General Assemblies of the IAU in the 1920s and 1930s. In 1921 he was elected chairman of the Executive Committee of the Astronomische Gesellschaft, a position he kept until 1930.

Rebsdorf, Simon O. "Bengt Strömgren: Growing up with Astronomy, 1908-1932."
*Journal for the History of Astronomy* 34 (2003): 171-199.

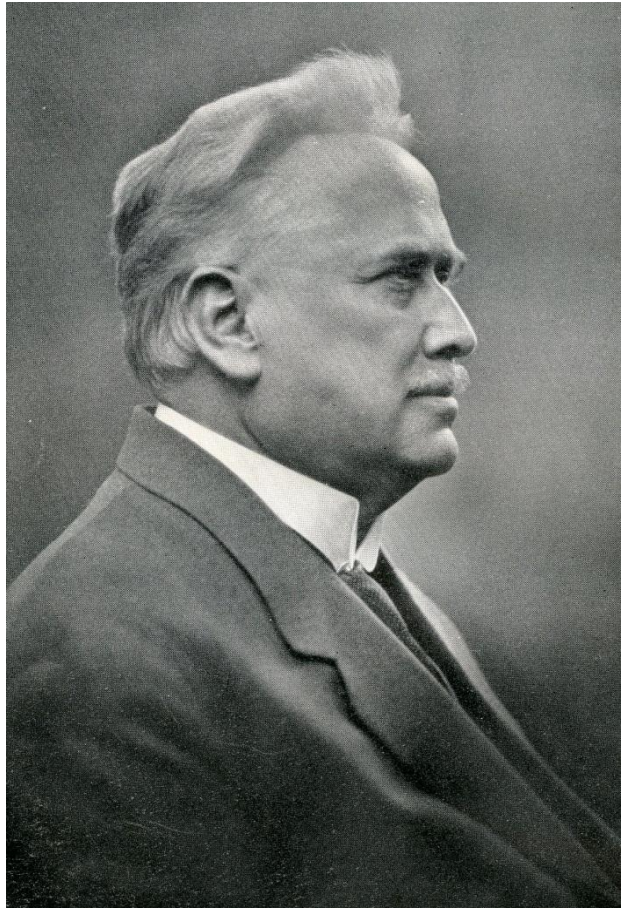

Svante Elis Strömgren